\documentstyle[emulateapj]{article}

\newcommand{\om}{\Omega_m}

\newcommand{\vs}{{v_{{\scriptscriptstyle 12}}}}

\newcommand{\se}{{\sigma_{{\scriptscriptstyle 8}}}} 
\newcommand{\be}{\begin{equation}}
\newcommand{\ee}{\end{equation}}

\newcommand{\lsim}{\stackrel{<}{\sim}}
\newcommand{\bea}{\begin{eqnarray}}
\newcommand{\eea}{\end{eqnarray}}
\newcommand{\bean}{\begin{eqnarray*}}
\newcommand{\eean}{\end{eqnarray*}}

\newcommand{\Mlu}{\,h^{-1}{\rm Mpc}}

\begin{document}

\submitted{The Astrophysical Journal, 558: L1-L4, 2001, September 1}

\title{ 
Gravity's smoking gun?} 
\author{ 
Enrique Gazta\~naga$^{1,2}$, Roman Juszkiewicz$^{3,4,5}$ \\
$^{1}$ INAOE, Astrofisica, Tonantzintla, 
Apdo Postal 216 y 51,  Puebla 7200, Mexico \\
$^{2}$ IEEC/CSIC, Edf. Nexus-201 - c/ Gran Capitan 2-4, 08034 Barcelona, Spain \\
$^{3}$ Physique Th{\'e}orique, Universit{\'e}
de Gen{\`e}ve, CH-1211 Gen{\`e}ve, Switzerland \\
$^{4}$ Institute of Astronomy, Zielona G{\'o}ra University,
  PL-62565 Zielona G{\'o}ra, Poland  \\
$^{5}$ Copernicus Astronomical Center, 00-716 Warsaw, Poland}


\begin{abstract} 
  
  We present a new constraint on the biased galaxy formation picture. 
  Gravitational instability theory predicts that the two-point mass density
  correlation function, $\xi(r)$, has an inflection point at the
  separation $r=r_o$, corresponding to the boundary between the linear and
  nonlinear regime of clustering, $\xi \simeq 1$.  We show how this feature
  can be used to constrain the biasing parameter, $b^2 \equiv \xi_g(r)/\xi(r)$
  on scales $r \simeq r_o$, where $\xi_g$ is the galaxy-galaxy correlation
  function, allowed to differ from $\xi$.  We apply our method 
  to real data: the $\xi_g(r)$, estimated from the
  {\sc APM} galaxy survey.  Our results suggest that the {\sc APM} galaxies
  trace the mass at separations $r \ga 5 \Mlu$, where $h$ is the Hubble
  constant in units of 100 km s$^{-1}$Mpc$^{-1}$. The present results agree 
  with earlier studies, based on comparing higher order correlations in 
  the {\sc APM} with weakly nonlinear perturbation theory. 
  Both approaches constrain
  the $b$ factor to be within $20\%$ of unity. If the existence of the feature
  we identified in the {\sc APM} $\xi_g(r)$ -- the inflection point near
  $\xi_g = 1$ -- is confirmed by more accurate surveys, we may have discovered
  gravity's smoking gun: the long awaited ``shoulder'' in $\xi$, predicted by
  Gott and Rees 25 years ago.

\end{abstract}

\keywords{Cosmology  -- large-scale structure of universe}

\section{Introduction}

The concept that galaxies may not be fair tracers of the mass distribution was
introduced in the early 1980s, in part in response to the observation that
galaxies of different morphological types have different spatial
distributions and that therefore they cannot all trace the mass 
(there are two excellent
reviews on the subject: Strauss \& Willick 1995 and Hamilton 1998). However,
there was also another reason: to ``satisfy the theoretical desire for a flat
universe'' (Davis et al. 1985, p.391).  More precisely, biasing was introduced
to reconcile the observations with the predictions of the Einstein-de Sitter
cold dark matter (CDM) dominated model.  At the time, it seemed that just a
simple rescaling of the overall clustering amplitude by setting $\xi_g(r) =
b^2\xi(r)$, where $b \approx 2$ might do the job (Davis et al. 1985).
However, very soon thereafter, it became clear that this is not enough: while
the unbiased ($b = 1$) $\xi(r)$ had too large an amplitude at small $r$, the
biased model did not have enough large-scale power to explain the observed
bulk motions (Vittorio et al. 1987, Ostriker 1993). 
A similar conclusion could be drawn form
comparison of the relative amplitude of clustering on large and small scales
(eg Maddox et al 1990).  The problem with the shape of $\xi(r)$ became
explicit when measurements of $\xi_g(r)$ showed that the optically selected
galaxies follow an almost perfect power law over nearly three orders of
magnitude in separation. This result disagrees with N-body simulations. The
standard ($\Omega_m = 1$) CDM model as well as its various modifications,
including $\Omega_m < 1$ and a possible non-zero cosmological constant, fail
to match the observed power law (see Fig 11-12 in Gazta\~naga 1995, Jenkins et
al. 1998; most of these problems were already diagnosed by Davis et al. 1985).
Two alternative ways out of this impasse were recently discussed by Rees
(1999) and Peebles (1999).
A possible response to the CDM crisis is to build a model where simple
phenomena, like the power-law behavior of $\xi_g$ are much more complicated
than they seem.  In particular, one can explore the possibility that the
emergence of large scale structure is not driven by gravity alone but by
``environmental cosmology'' -- a complex mixture of gravity, star formation
and dissipative hydrodynamics (Rees 1999). A phenomenological formalism,
appropriate for this approach was recently proposed Dekel \& Lahav (1999).  
 There are also some recent analytical models,
based on halo profiles and halo-halo clustering (Seljak 2000, Scoccimarro
etal 2001). An
obvious alternative to environmental cosmology was recently discussed by
Peebles (1999), who pointed out that ``as Kuhn has taught us, complex
interpretations of simple phenomena have been known to be precursors of
paradigm shifts''.
Instead, one can explore a simpler
option, that galaxies trace the mass distribution, at least for local (low
redshift), optically selected galaxies with a broad magnitude sampling.  This
approach rests on the idea that no matter how or where galaxies form, they
must eventually fall into the dominant gravitational wells and therefore trace
the underlying mass distribution (see Peebles 1980, hereafter LSS; Fry 1996).

The absence of biasing on large (weakly nonlinear) scales agrees well with
other observational evidence.  The measurements of the two-, three- and
four-point connected moments of the density field in the {\sc APM} catalog
provide support for the hypothesis that galaxies trace the mass and also
that the
large-scale structure we observe today grew out of small-amplitude, Gaussian
density fluctuations in an expanding, self-gravitating non-relativistic gas.
Indeed, the theoretical predictions for the first few connected moments, based
on this hypothesis (Juszkiewicz et al. 1993, Bernardeau 1994) are in good
agreement with the APM measurements (Gazta\~naga 1994, Gazta\~naga \& Frieman
1994, Frieman \& Gazta\~naga 1999, and references therein). The current
precision of this higher order correlation test is $20\%$ and expected to
improve with future data. The absence of biasing is also suggested by the most
recent measurements of the mean relative pairwise velocity of galaxies
(Juszkiewicz et al. 2000).

In this {\it Letter} we propose a new test of the biasing hypothesis.
Our test is based on the behavior of $\xi(r)$ at the $\xi =1$
nonlinearity boundary. 
We describe our theoretical model in the next section. It is
checked against N-body simulations in \S3. It is then applied to
the {\sc APM} survey in \S4. Finally, in \S5 we discuss our results.

\section{The inflection point}

In the gravitational instability theory, newly forming mass clumps are
generally expected to collapse before relaxing to virial equilibrium.  If this
were so, the relative mean 
velocity of particle pairs $|\vs(r)|$ would have to be
larger than the Hubble velocity $Hr$ to make $\vs(r) + Hr$ negative. As a
consequence of the pair-conservation equation (LSS), 
the slope $d \ln \xi(r) /d \ln r \equiv - \,
\gamma(r)$ must rapidly decrease with decreasing separation
near the boundary of nonlinearity, i.e. when $r \lsim r_o$. This
effect was recognized long ago by Gott \& Rees (1975). When the expected
``shoulder'' was not found in the observed galaxy-galaxy correlation function,
Davis \& Peebles (1977) introduced the so-called previrialization conjecture
as a way of reducing the size of the jump in $\gamma(r)$ (the conjecture
involves non-radial motions within the collapsing clump; see the discussion in
LSS, \S 71 and Peebles 1993, pp. 535 - 541; see also Villumsen \& Davis 1986;
{\L}okas et al. 1996 and 
Scoccimarro \& Frieman 1996).  Later observational work showed a shoulder
in $\xi$ in several redshift and angular catalogs, which was
also interpreted as evidence for the boundary between linear and
nonlinear gravitational clustering 
(see the review by Guzzo 1997 and references therein).

Quarter a century later the precision of N-body simulations as
well as the quality of the observational data have improved dramatically
enough to justify a reexamination of the problem.  The actual shape of the
correlation function near $\xi = 1$ can be investigated with high resolution
N-body simulations like those run by the Virgo Consortium (Jenkins et al.
1998).  As shown in Juszkiewicz et al. (1999, hereafter {\sc JSD}),
in all four of the Virgo models (SCDM, $\Lambda$CDM, OCDM, $\tau$CDM)
the slope of $\xi(r)$ exhibits a
striking feature. Instead of a shoulder, or a simple discontinuity in
$\gamma(r)$, however, $\xi(r)$ has an inflection point,
$d^2\xi(r)/dr^2 = 0$
which occurs at a uniquely defined separation $r = r_*$.  At this separation,
the logarithmic slope of $\xi$ reaches a local maximum, $\, d \ln\xi/d\ln r =
- \gamma_*$.  In all models JSD investigated, the inflection point
indeed appears near the transition $\xi = 1$, as expected by the earlier
speculations, involving the ``shoulder'' in $\xi$.  The separation 
$r_*$ is almost
identical with the scale of nonlinearity:
\begin{equation}
r_* \; \approx \; r_o \;, \qquad \xi(r_o) \; \equiv \; 1 \; .
\label{test}
\end{equation}
More precisely, a comparison of Figure 1 in JSD with
Figure 8 in Jenkins et al. (1998) gives 
\begin{equation}
|r_o - r_*| \; < \;0.1 \,r_o
\label{precise_test}
\end{equation}
for all four considered models. 
Moreover, for all models, studied by JSD, the
$- \gamma$ vs. $r$ dependence can be described as an S-shaped curve, with a
maximum at $r = r_* \approx r_o$, and a minimum at a smaller separation. 
The depth of the minimum in $-\gamma(r)$ increases with increasing normalization
parameter, $\sigma_8$ --
the final linear rms mass density contrast, 
measured in spheres of a radius of 8 $h^{-1}$Mpc 
(see Fig.8 in JSD).
If the relation (\ref{test}) is indeed a general property of gravitational
clustering, it can be used as a test of biasing as follows.  Suppose the
biasing factor is significantly greater than
unity: $b \gg 1$. Then $\xi_g \gg \xi$ and the relation (\ref{test}) will
break down.  For power-law galaxy correlation function, $\xi_g (r) =
(r_{og}/r)^{\gamma} = b^2\xi(r)$, and instead of equation 
(\ref{test}) we will have
$r_* \; \approx \; r_{og} \, b^{-2/\gamma}$.
Since the observed slope is $\gamma \approx 1.8$, for $b = 2$, the shoulder in
the correlation function should appear at a separation smaller than a half of
the $r_{og}$ parameter! 
The technique we propose is unable to
constrain more baroque biasing models with a large number of free parameters.
However, the predictive power of such models is questionable
and one may ask: are they falsifiable at all?

\section{APM-like N-body simulations}

In this section, we compare the JSD analytic and N-body results with a
different set of P$^3$M simulations.  Instead of the family of CDM models,
considered by JSD, we use {\sc APM}-like initial conditions, which have
Gaussian initial conditions with an initial power spectrum designed to evolve
into a final spectrum, matching the APM measurements, under the additional
assumption of no bias and $\om=1$.  We use simulations with identical APM-like
spectrum, with $\Lambda = 0$ and two different values of the density
parameter: $\om = 1$ and 0.3.  The box size is $600 ~h^{-1}$ Mpc (or $300
~h^{-1}$ Mpc) with $200^3$ (or $100^3$) dark matter particles with $\se=0.85$
(for more details see Baugh \& Gazta\~naga 1996). 

\begin{figure*}
\centerline{\epsfysize=8truecm
\epsfbox{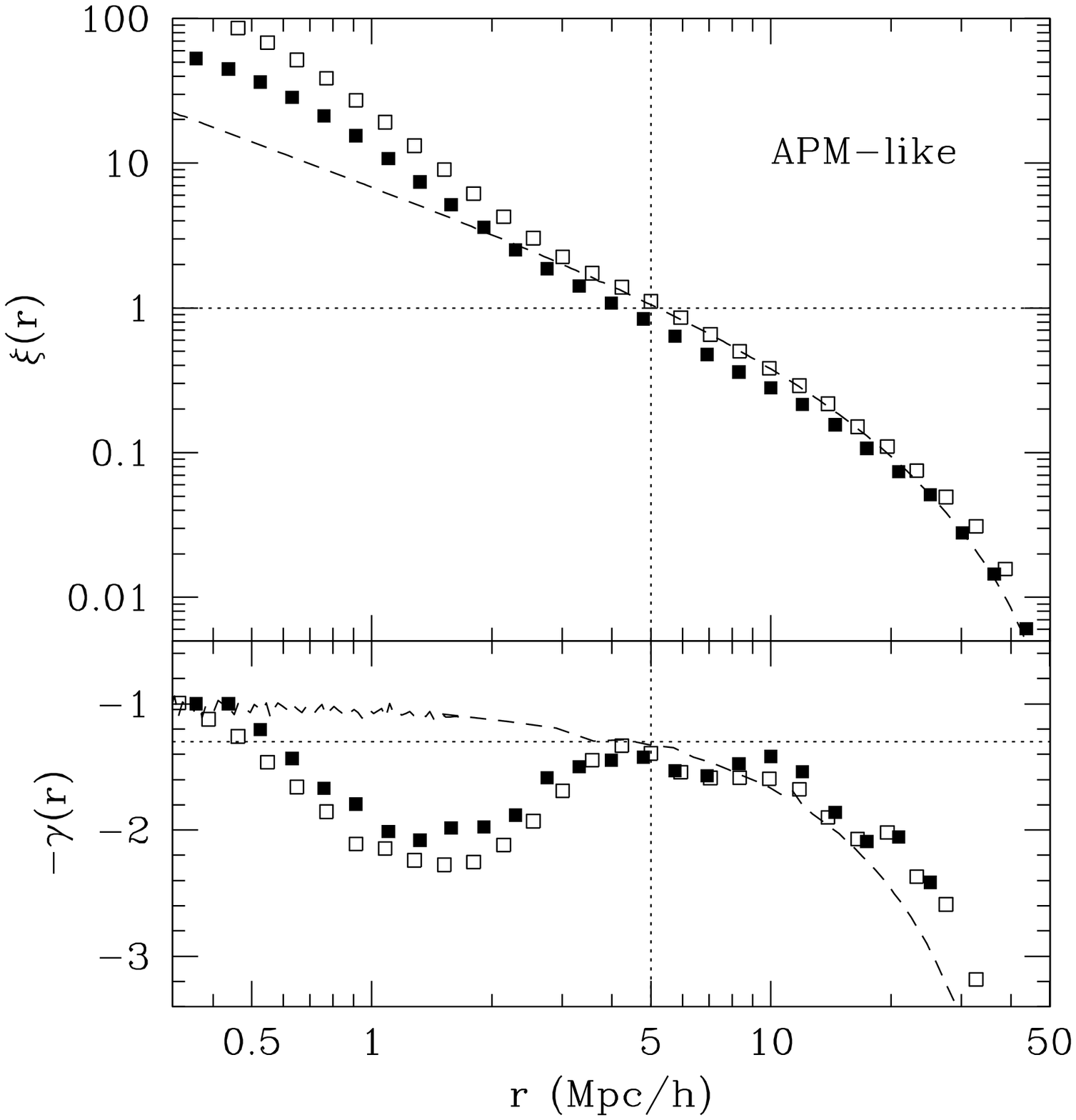}\epsfysize=8truecm
\epsfbox{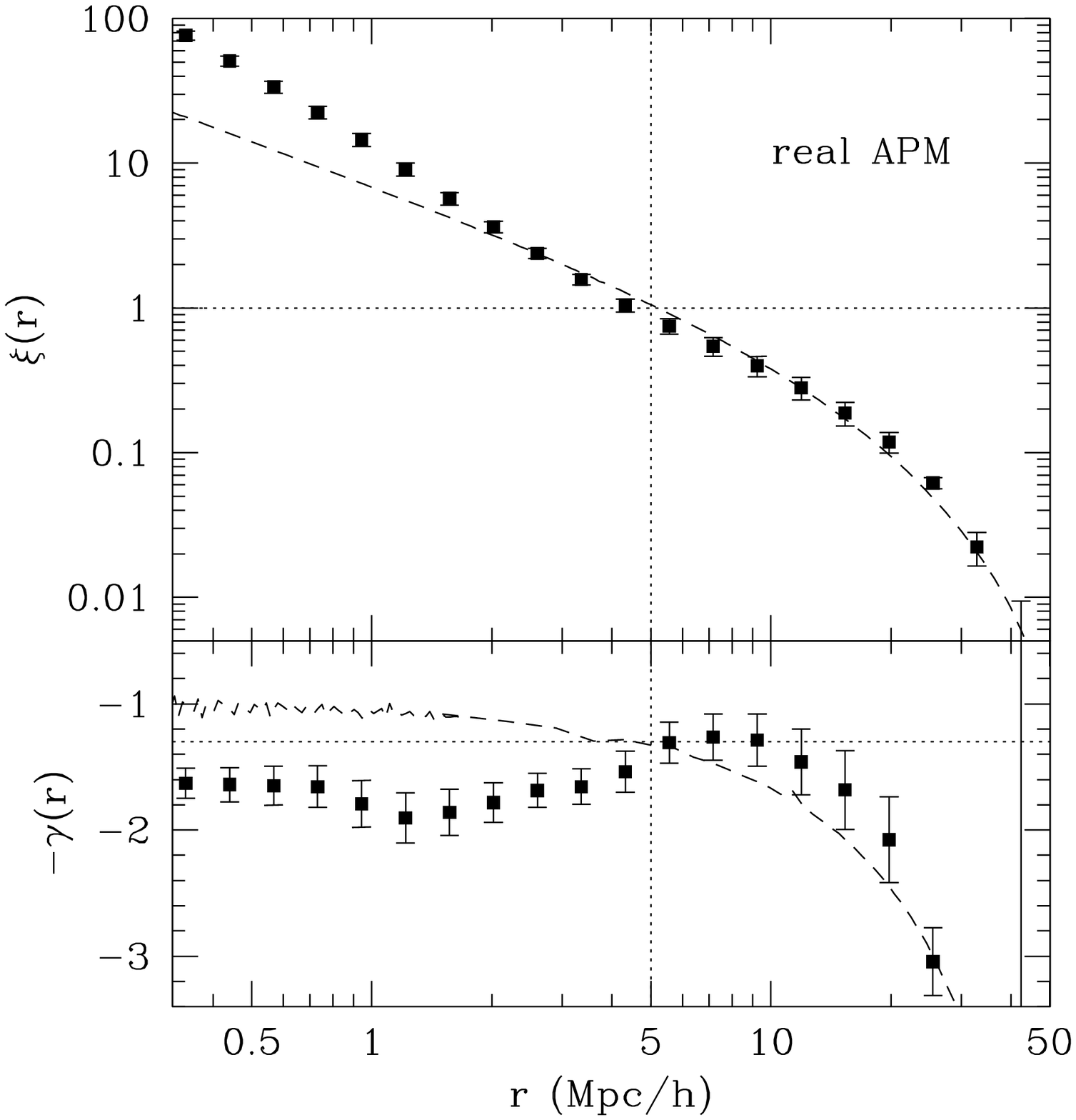}}
\caption{ 
    {\sl Top left panel}: linear $\xi(r)$ (dashed line) and the measured
  nonlinear $\xi(r)$, obtained from the APM-like simulations with
  $\Omega_m=0.3$ (open squares) and $\Omega_m=1.0$ (full squares). 
{\sl Bottom
  left panel:} Corresponding logarithmic slope, $\, -\gamma(r) = d \ln
  \xi /d\ln r$ for each of the three curves from the top panel. The right 
panels show similar results for the spatial $\xi(r)$ estimated for
  {\sc APM} galaxies ({\sl symbols with errorbars}), compared to the same 
linear theory
  {\sc APM-}like model ({\sl dashed line}).   The vertical
  dotted lines shows the scale $r_o$, defined by the condition $\xi (r_o) = 1$
  (top) and the scale $r_*$, at which the nonlinear $\gamma(r)$ curve crosses
  the linear one (bottom). }
\label{x2slapm}
\end{figure*}

The evolved, nonlinear correlation functions, measured from simulations are
shown in Figure \ref{x2slapm} (top left panel).  
 \notetoeditor{The Figures in the two postscript files
 should appear side-by-side, double column, in print} 
The full squares correspond to
the $\om = 1$ model, while the open squares represent $\om = 0.3$. For
comparison, we show the linear correlation function (dashed line).
Nonlinear effects are more pronounced in the low
density model. Note however, that although the correlation functions differ
significantly in amplitude at separations $r < 2 h^{-1}$Mpc, their 
slopes $\gamma(r)$ are almost indistinguishable.

The particle resolution (the Nyquist wavelength $\propto N^{-1/3}$) of the
simulations used here is significantly lower than the resolution of Virgo
simulations, and the noise in the measured $\xi(r)$ is further amplified by
differentiating over $r$. As a result, determining the position of the
inflection point $r_*$ directly from the $\gamma(r)$ curve alone is more
difficult. A noise-resistant, alternative definition of $r_*$,
suggested by JSD is to identify $r_*$ with the separation at which the
linear $\gamma(r)$, derived from initial conditions, crosses the
nonlinear $\gamma(r)$ curve, measured from the simulations. This approach 
is indeed effective for our APM-like simulation (see Fig.1), but
it can not be applied to the true APM data since the we know only
the nonlinear $\gamma(r)$.  The observed $-\gamma(r)$ curve does resemble
the S-shaped $-\gamma(r)$ from the VIRGO simulations; the main
difference is that the peak as well as the trough are broad
and fuzzy rather than narrow and sharp as in the simulations. 
To deal with this problem, we used the
following prescription. For a discrete set of
measurements ${r_i,\gamma(r_i)}$, $i = 1,2,\ldots$ starting with
some separation $r_1$ rightward of the peak, and moving to smaller
separations, we compare consecutive values of $-\gamma(r_i)$ and
$-\gamma(r_{i+1})$. We then identify $r_*$ with the largest $r_i$,
for which $-\gamma(r_i)$ drops below its maximum value for the next
few points (this is to avoid picking up local maxima due to fluctuations,
see Fig.1). Obviously, the above prescription would also pick $r_*$ in high
resolution simulations. When applied to the APM-like simulations,
this method gives $r_*\simeq 4-5\, h^{-1}$ Mpc. The same simulations also
give $r_o \simeq 5\, h^{-1}$ Mpc,
in excellent agreement with equation (\ref{precise_test}), which we will
consider as a measure of a systematic error, introduced by the
theoretical model we use. From these simulations we conclude that
equality between $r_*$ and $r_o$ can probably be considered as a generic
outcome of gravitational dynamical evolution in a model where galaxies trace
the mass and the initial slope, $\, d\ln\xi/d\ln r$, is a smooth decreasing
function of the separation $r$ (as expected in hierarchical
clustering models, see e.g. LSS).

\section{Comparison with observations}

The measurements of $\xi_g(r)$, and $\, \gamma(r) \equiv - d\ln
\xi_g/d\ln r$, obtained from the angular correlations of galaxy pairs
in the {\sc APM} catalog (Baugh 1996), are plotted in Figure
\ref{x2slapm}.  Errorbars correspond to the dispersion in $\gamma(r)$
from 4 APM strips of about 30$\times$60 degrees each, eg 200$\times$400 
$\Mlu$ at the mean depth of the APM. 
At scales of $r \simeq 5 \Mlu \ll 200 \Mlu$ most of the covariance
between consecutive bins in $\xi_g$ 
is due to large scale density fluctuations
that shift the mean density between strips. This introduces covariance in
the amplitude of $\xi_g(r)$ (the whole $\xi_g$ curve is shifted vertically
from strip to strip, see eg Fig.2 in Baugh 1996), 
but does not affect much its shape  $\gamma(r)$.
Thus, we consider our errors in  $\gamma(r)$ as independent. 
The top left panel shows the two-point function (points
with error bars), and the linear theory curve, described in \S 3
(dashed line). The intersection of the two perpendicular dotted
lines marks the point $(\xi_g, r) = (1,  r_{og})$.  The
bottom panel of Figure \ref{x2slapm} shows the {\sc APM} $\gamma(r)$
as a function of the pair separation $r$. Note the remarkable
similarity between the empirical data and the characteristic peak in
the $-\gamma(r)$ found in the simulations (compare the left and right  bottom
panels in Fig.1; see also Fig.1 in JSD).  The
intersection of the two mutually perpendicular, dotted lines in the bottom
panel of Figure \ref{x2slapm} marks the result of applying
our prescription for estimating $r_*$ to the APM data. 
The slope $-\gamma$ drops down from its maximum value at
the separation $r \simeq 5\Mlu$, and to first approximation this scale
could be identified with $r_*$. 
Taking into account the error bars in Figure
\ref{x2slapm} we obtain $ r_* \simeq (5.5 \pm 1.5)~\Mlu$,
$ r_{og} \simeq (4.5 \pm 0.5)~\Mlu $ and $\gamma_* \simeq -1.4$.
From these measurements one can
estimate $b$ in the linear bias model
\be
b^2= (r_{og}/r_*)^{\gamma_*}. 
\ee
This expression for $b^2$ is also 
exactly valid for a nonlinear scale-dependent bias at $b(r_o)$ 
as far as $r_* \simeq r_0$:
$b^2(r_o) \equiv \xi_g(r_o)/\xi(r_o) \equiv (r_{og}/r_o)^{\gamma(r_o)} = 
(r_{og}/r_*)^{\gamma_*}$. This is true even if the slopes of the galaxy and matter
correlations are different at $r=r_0$. We find
 
\be 
b(r_o) \simeq \; 1.15 \pm 0.23 ~(\pm 0.11) \;
\label{b-constraint}
\ee
at one-sigma level in the errors. The error in parenthesis
corresponds to the systematic uncertainty in Eq.\ref{precise_test}.

\section{Discussion}

Recently, Hamilton and Tegmark (2000) found no evidence of an inflection
at the linear-nonlinear transition scale in $\xi_g$, estimated from the
PSCz survey.  The origin of this difference with our APM results is not
clear, since a recent study of the three-point correlations in the PSCz
catalog (Feldman et al. 2001) leaves little room for biasing, providing
constraints on $b$ similar to those obtained here and to those obtained
earlier from the measurements of three-point correlations in the APM
catalog.  If the APM inflection as well as its absence in the PSCz are
both real phenomena, it then means that APM galaxies are less biased than
PSCz (IRAS) galaxies.  Apart from constraining the linear bias, Feldman et
al. have also measured a small but statistically significant second-order
biasing parameter, consistent with the observation that infrared-selected
(IRAS) galaxies avoid high density cores of clusters. We plan a systematic
study of this effect in near future.

We are impressed how well the shape of $\gamma(r)$ in the APM observations
resembles gravity's ``shoulder''.  This feature is a robust result from the
APM catalogue and can be seen directly in the angular 2-point function and in
the recovered shape of the power spectrum (Baugh \& Efstathiou 1993).
Numerical simulations (Gazta\~naga \& Baugh 1996) show that this is not an
artifact of the de-projection. It is difficult to imagine how such an
agreement could happen by a mere coincidence, which would have to be the case
if $\xi_g$ is unrelated to $\xi$.  Our results are by no means final, they are
also less rigorous than one could wish because we are limited by the accuracy
of the present observational data.  New generation of catalogs promise an
improvement on this front in the near future (for an excellent collection of
reports on the state of the art in this field, see Colombi et al. 1998).

{\it Acknowledgments.}  We thank Carlton Baugh for providing his {\sc
  APM}-like simulations and his estimate of $\xi_g(r)$, based on the {\sc APM}
survey. RJ thanks Ruth Durrer for important discussions regarding the
inflection point in $\xi(r)$. We also acknowledge support from a collaborative
grant between the Polish Academy of Sciences and the Spanish CSIC, grants from
the Polish Government (KBN grant No.2P03D01719), from the Swiss Tomalla
Foundation, and from IEEC/CSIC and DGES(MEC) (Spain), project PB96-0925.






\begin{thebibliography}{ZZZZZZZZZZZ1999}


\def\refe {\par \hangindent=.7cm \hangafter=1 \noindent}
\def\apj {ApJ}
\def\na {Nature,}
\def\aap {A\&A}
\def\apjs{ApJS}
\def\mn {MNRAS}

\bibitem [Baugh & Efstathiou 1993]{be93}
Baugh, C.M., \& Efsthathiou, G., 1993, \mn, 265, 145

\bibitem [Baugh 1996]{b96}
Baugh, C.M., 1996, \mn, 282, 1413

\bibitem [Baugh \& Gazta\~naga 1996]{bg96}
Baugh, C.M., \& Gazta\~naga, E., 1996, \mn, 280, 37

 
\bibitem [Bernardeau 19994]{b94b}  Bernardeau, F., 1994, \apj, 433, 1

\bibitem [Colombi et al. 1998]{stephane} Colombi, S.,
Mellier, Y., \& Raban, B., eds., 1998, Wide Field Surveys 
in Cosmology, Editions Frontieres, Paris.

\bibitem[Davis \& Peebles, 1977]{davis77} Davis, M. \& Peebles,
P.J.E., 1977 \apjs, 34, 425


\bibitem [Davis et al. 1985] 
{gang_of_4} Davis, M., Efstathiou, E., Frenk, C.S., White, C.D.M.,
1985, \apj, 292, 371

\bibitem[Dekel \& Lahav 1999] 
{ofer} Dekel, A., \& Lahav, O., 1999, \apj, 520, 24

\bibitem[Feldman 2001]{feldman}
Feldman, H.A., Frieman, J.A., Fry, J.N., Scoccimarro, R.,
2001, Phys Rev D. in press, astro-ph/0010205 


\bibitem[Frieman \& Gazta\~naga 1999]{gf99}  Frieman, J.A.,
Gazta\~naga, E., 1999, \apj,  521, L83 

\bibitem[Fry 1996]{jf96} 
Fry, J., 1996, \apj, 461, L65

\bibitem[Gazta\~naga 1994]{gaz94} Gazta\~naga, E., 1994, \mn,
268, 913 

\bibitem[Gazta\~naga 1995] {g95} 
Gazta\~naga, E. 1995, \apj, 454, 561

\bibitem[Gazta\~naga \& Baugh 1998]{gb98} Gazta\~naga, E. \& Baugh, C.M.
1998, \mn, 294, 229 

\bibitem[Gazta\~naga \& Frieman 1994]{gf94} Gazta\~naga, E.,
 Frieman, J.A., 1994, \apj, 437, L13 


\bibitem [Gott \& Rees 1975] {rich} 
Gott, J.R., \& Rees, M.J., 1975, \aap, 45, 365

\bibitem [Guzzo 1997] 
{gu97} Guzzo, L., 1997, New Astronomy, 2, 517


\bibitem[Hamilton \& Tegmark 2000]{ht2000} Hamilton, A.J.S.,
Tegmark, M., 2000, astro-ph/0008392


\bibitem[Hamilton 1998]{andrew-review} Hamilton, A.J.S., 1998,
The Evolving Universe, Kluwer, Dordrecht,
p. 185

\bibitem[Jenkins et al. 1998]{virgo} 
Jenkins, A. et al. (The Virgo Consortium),
1998, \apj, 499, 20

\bibitem[Juszkiewicz et al. 1993]{rj93}
Juszkiewicz, R., Bouchet, F.R., \& Colombi, S., 1993, \apj 412, L9  

\bibitem[Juszkiewicz et al. 1999]{rj99}
Juszkiewicz, R., Springel, V., \& Durrer, R., 1999,
\apj, 518, L25 

\bibitem[Juszkiewicz et al. 2000]{rj00}
Juszkiewicz, R., Ferreira, P.G., Feldman, H.A., Jaffe, A.H., \& Davis, M., 2000,
Science, 287, 109


\bibitem[{\L}okas et al. 1996]{ewa}
{\L}okas, E., Juszkiewicz, R., Bouchet, F.R., \& Hivon, E., 1996,
\apj, 467, 1

\bibitem[Maddox et al. 1990]{mad} 
Maddox, S.J., Efstathiou, G., Sutherland, W.J. \& Loveday,
J., 1990, \mn, 242, 43P

\bibitem[ostriker]{ost}
Ostriker, J.P. 1993, Annual Review of A\&A, 31, 689.

\bibitem[Peebles 1980]{peebles80} Peebles, P.J.E., 1980, The
Large--Scale Structure of the Universe,
Princeton University Press, Princeton (LSS)

\bibitem[Peebles 1993]{jim93} Peebles, P.J.E., 1993, Principles
of Physical Cosmology, Princeton University Press

\bibitem[Peebles 1999]{jim99} Peebles, P.J.E., 1999, 
in Clustering at High Redshift, eds. A. Mazure \& O. Le Fevre 
astro-ph/9910234

\bibitem[Rees 1999]{rees} Rees, M.J.,  1999, preprint,
astro-ph/9912373

\bibitem[Scoccimarro \& Frieman 1996]{rs96} Scoccimarro, R.,
\& Frieman, J., 1996, \apj, 473, 620

\bibitem[Scoccimarro, Sheth, Hui, \& Jain(2001)]{2001ApJ...546...20S}
Scoccimarro, R., Sheth, R.K., Hui, L., Jain, B. 2001, \apj, 546,
20 

\bibitem[Seljak(2000)]{2000MNRAS.318..203S} Seljak, U.~; 2000, \mnras,
318, 203 

\bibitem[(Strauss \& Willick  1995)]{strausswillick95}Strauss, M., 
 Willick, J., 
1995, Physics Reports, 26, 271

\bibitem [Villumsen \& Davis 1986] {jens} 
Villumsen, J., \& Davis, M., 1986, \apj 308, 499

\bibitem [Vittorio et al. 1987] {vjd} 
Vittorio, N., Juszkiewicz, R., \& Davis, M,
1986, Nature, 323, 132 

\end{thebibliography}
\end{document}